\documentclass[%
reprint,
 superscriptaddress,
 amsmath,amssymb,
 aps,
longbibliography]{revtex4-1}

\usepackage{graphicx}
\usepackage{subfigure}
\usepackage{dcolumn}
\usepackage{bm}
\usepackage{epstopdf}
\usepackage{xcolor}
\usepackage{hyperref}

\begin{document}


\title{Low-Energy Scattering Properties of Ground-State and Excited-State Positronium Collisions }

\author{Michael D. Higgins}%
\affiliation{Department of Physics and Astronomy, Purdue University, West Lafayette, Indiana 47907, USA}
\email{higgin45@purdue.edu}
\author{Kevin M. Daily}
\affiliation{Department of Physics and Astronomy, Purdue University, West Lafayette, Indiana 47907, USA}
\email{ke.daily@gmail.com}
\author{Chris H. Greene}
\affiliation{Department of Physics and Astronomy, Purdue University, West Lafayette, Indiana 47907, USA}
\affiliation{Purdue Quantum Center, Purdue University, West Lafayette, Indiana 47907, USA}
\email{chgreene@purdue.edu}

\date{\today}

\begin{abstract}
Low-energy elastic and inelastic scattering in the Ps(1$s$)-Ps(2$s$) channel is treated in a four-body hyperspherical coordinate calculation. Adiabatic potentials are calculated for triplet-triplet, singlet-singlet, and singlet-triplet spin symmetries in the spin representation of coupled electrons and coupled positrons, with total angular momentum $L=0$ and parity equal to $+1$. The s-wave scattering lengths for the asymptotic Ps(1$s$)-Ps(2$s$) channel are calculated for each spin configuration. Results obtained for the s-wave scattering lengths are $a_{\mathrm{TT}}=$~$7.3(2)a_0-i0.02(1)a_0$, $a_{\mathrm{SS}}=$~$13.2(2)a_0-i0.9(2)a_0$, and $a_{\mathrm{ST}}=$~$9.7(2)a_0$ for each spin configuration. Spin recoupling is implemented to extract the scattering lengths for collisions of Ps in different spin configurations through properly symmetrized unitary transformations. Calculations of experimentally relevant scattering lengths and cross-sections are carried-out for Ps atoms initially prepared in different uncoupled spin states.
\end{abstract}

\pacs{Valid PACS appear here}
\maketitle

\section{\label{sec:level1}Introduction}

The possibility of creating a gamma-photon laser from a Bose-Einstein Condensate(BEC) of spin-polarized positronium(Ps) is of considerable interest \cite{Cassidy2018,Cassidy2005,PlatzmanMills,Mills2002,Avetissian2014,Wang2014}. The main mechanism behind such a laser is the annihilation of the electron and the positron. This annihilation process would release photons with a combined energy of $\sim1.02\ MeV$ per particle-antiparticle pair, corresponding to wavelengths on the order of pico-meters. Since particles and antiparticles annihilate, the Ps atom has a finite lifetime. As is well known, the mean lifetime for the Ps atom depends on its spin state. Triplet Ps has a mean lifetime of $142\ ns$ and $1.136\ \mu s$ for $1s$ and $2s$ states, respectively, while singlet Ps has a mean lifetime of $0.125\ ns$ and $1\ ns$ for the $1s$ and $2s$ states, respectively \cite{Cassidy2018,OrePowell,dirac_1930}. 

To realize a Ps BEC, high density ensembles of spin-polarized triplet Ps atoms should be created and thoroughly studied. Recently, experimental efforts have been underway to produce Ps ensembles using positron beams, even leading to the creation of the Ps molecule, $\mathrm{Ps_{2}}$ \cite{CassidyMills2007,CassidyMeligneMills2010,Cassidy2005, CassidyDengMills2007}. Positronium densities high enough for BEC creation have yet to be obtained, but current systems are good systems for studying Ps--Ps scattering. One way of measuring Ps densities that is of interest, and pertains to this study, is via cold collision frequency(clock-shift) measurements, which has been carried out for hydrogen ensembles.

Significant theoretical and experimental progress has been made to produce and study the BEC of spin-polarized hydrogen \cite{Killian1998,Stwalley,Killian2000,KillianThesis}. Methods to experimentally probe the density of an atomic gas have been explored \cite{Killian1998,KillianThesis,FriedKillian1998,Jamieson1996} from measurements of the clock-shift frequency of ground-state (n=1) and first excited-state (n=2) spin-polarized hydrogen. The clock-shift is a shift in frequency of line transitions between energy levels due to atomic collisions. The clock-shift frequency is proportional to the density of the gas and the difference between the atom-atom s-wave scattering lengths in the ground-excited and ground-ground states (see Eqs. 1 and 2 of \cite{FriedKillian1998}).

To measure densities of Ps gases via the clock-shift, scattering properties of Ps-Ps collisions should be understood. Extensive numerical studies have been performed to determine the 1$s$--1$s$ triplet-triplet(TT) and singlet-singlet(SS) scattering lengths for ground-state Ps-Ps collisions \cite{Ivanov2001,Ivanov,ADHIKARI2002308,CHAKRABORTY2004112,Oda2001,Shumway2001,e_+e_-Daily}. However, collisions of ground-state Ps and excited-state Ps have not been studied in much detail, even though it is important because it would provide researchers with a key tool to probe Ps densities. This recognition forms the motivation for the present study. A study of Ps(1$s$)-Ps(2$s$) collisions is conducted in order to provide quantitative estimates of the 1$s$-2$s$ $s$-wave scattering lengths for different spin orientations. 

This paper is organized in the following way. First, Section \ref{sec:background} describes the general set-up of the system under study, the calculated adiabatic potential curves, and the clock-shift. In Section \ref{sec:numanalysis}, calculations of the scattering lengths are presented for the TT, SS, and singlet-triplet(ST) spin configurations as well as a brief error analysis. Then Section \ref{sec:recoupling} discusses spin recoupling from the $e^-e^-$ and $e^+e^+$ spin representation to the physically relevant Ps spin representation. Section \ref{sec:crosssection} discusses partial cross-sections for different uncoupled spin states of Ps collisions.

\section{Theoretical Background}
\label{sec:background}


This section briefly summarizes the theory provided in Daily et al. \cite{e_+e_-Daily} to describe the problem being solved. Also, a discussion of the calculated adiabatic potential curves used to obtain scattering properties are discussed in detail.
\subsection{The Hamiltonian of the Ps-Ps System}
The system being studied is the collision between two positronium atoms. This system consists of two electrons and two positrons and the Hamiltonian of such a system can be written in Cartesian coordinates as \cite{e_+e_-Daily,DailyAsym},
\begin{equation}
    H=-\frac{1}{2}\sum_{i=1}^{4}\nabla^2_{\vec{r_{i}}}+\sum_{i \neq j}\frac{q_i q_j}{\left|\vec{r_i}-\vec{r_j}\right|}
    \label{eq:catresianham}
\end{equation}
where $\vec{r_i}$ is the position vector of the $i_{th}$ particle, and $q_i$ is its charge. In this paper, particles $1$ and $2$ represent positrons and particles $3$ and $4$ represent electrons. This Hamiltonian neglects annihilation and other relativistic effects such as spin interactions. The spin-spin interaction between two Ps atoms is zero due to the absence of a magnetic moment for triplet Ps \cite{Cassidy2018}. The spin-orbit interaction between the two Ps atoms in $s$-orbitals is zero for total orbital angular momentum quantum number  $L=0$, considered in this study.

For a description of the diverse ways this system can fragment, the four-body Hamiltonian is conveniently represented in hyperspherical coordinates \cite{,e_+e_-Daily,DailyAsym,fourfermionhyper}:
\begin{equation}
    H=H_{rel}+H_{C.M.}
    \label{eq:Hgeneral}
\end{equation}
where $H_{C.M.}$ is the Hamiltonian of the center of mass of the positronium system and $H_{rel}$ is the Hamiltonian describing the relative motion with respect to the center-of-mass frame,
\begin{equation}
    H_{rel}=-\frac{1}{2\mu}\frac{1}{R^8}\frac{\partial}{\partial R}(R^8\frac{\partial}{\partial R})+T_{\Omega}+V_{int}(R,\Omega)
    \label{eq:Hrel}
\end{equation}
where $R$ is the hyperradial coordinate, $\Omega$ represents the hyperangular coordinates, $\mu$ is the hyperspherical reduced mass, $T_{\Omega}$ is the hyperangular kinetic energy, and $V_{int}(R,\Omega)$ is the interaction potential. For relatively straightforward definitions of $R$, $\mu$, $T_{\Omega}$ and $V_{int}(R,\Omega)$, see \cite{DailyAsym}. 

The Hamiltonian is quasi-separable in the hyperradial and hyperangular coordinates, which leads to the standard ansatz for the solutions of Eq. \ref{eq:Hrel},
\begin{equation}
    \psi_{E}(R,\Omega)=\frac{1}{R^4}\sum_{\nu}F_{E,\nu}(R)\phi_{\nu}(R,\Omega)
    \label{eq:ansatz}
\end{equation}
where $\psi_{E}$ is represented as an eigenfunction expansion in the orthonormal basis set $\phi_{\nu}(R,\Omega)$ at fixed $R$ of the adiabatic Hamiltonian $H_{ad}(R,\Omega)=\frac{\Lambda^2+12}{2 \mu R^2}+V_{int}(R,\Omega)$ (see \cite{e_+e_-Daily,DailyAsym} for details). This leads to the following eigenvalue problem represented in Eq. \ref{eq:Had}. 
\begin{equation}
    \label{eq:Had}
    H_{ad}(R,\Omega)\phi_{\nu}(R,\Omega)=U_{\nu}(R)\phi_{\nu}(R,\Omega)
\end{equation}
The eigenvalues $U_{\nu}(R)$ of Eq. \ref{eq:Had} represent adiabatic potential curves that describe the interactions between the electrons and positrons as a function of the hyperradius, which is assumed to vary slowly compared to electronic motion. The radial functions $F_{E,\nu}(R)$ are obtained by solving the coupled radial equations $\left<\phi_{\mathrm{\nu^\prime}}\right|H-E\left|\psi_{E}\right\rangle=0$, which leads to Eq. \ref{eq:coupled},
\begin{multline}
    \label{eq:coupled}
    \left(-\frac{1}{2\mu}\frac{\partial^2}{\partial{R^2}}+U_{\mathrm{\nu}}\left(R\right)-E\right)F_{\mathrm{E,\nu}}\left(R\right)\\
    -\frac{1}{2\mu}\sum_{\nu^\prime}\left(2P_{\nu\nu^\prime}(R)\frac{\partial}{\partial{R}}+Q_{\nu\nu^\prime}(R)\right)F_{E,\nu^\prime}(R)=0
\end{multline}
where $P_{\nu\nu^\prime}(R)$ and $Q_{\nu\nu^\prime}(R)$ are non-adiabatic coupling matrices that couple channels with standard definitions given by Eqs. 10 and 11 in \cite{e_+e_-Daily}.

\subsection{Adiabatic and Diabatic-like Potentials}
\begin{figure}[!ht]
    \centering
    \subfigure[]{\includegraphics[width=1\columnwidth]{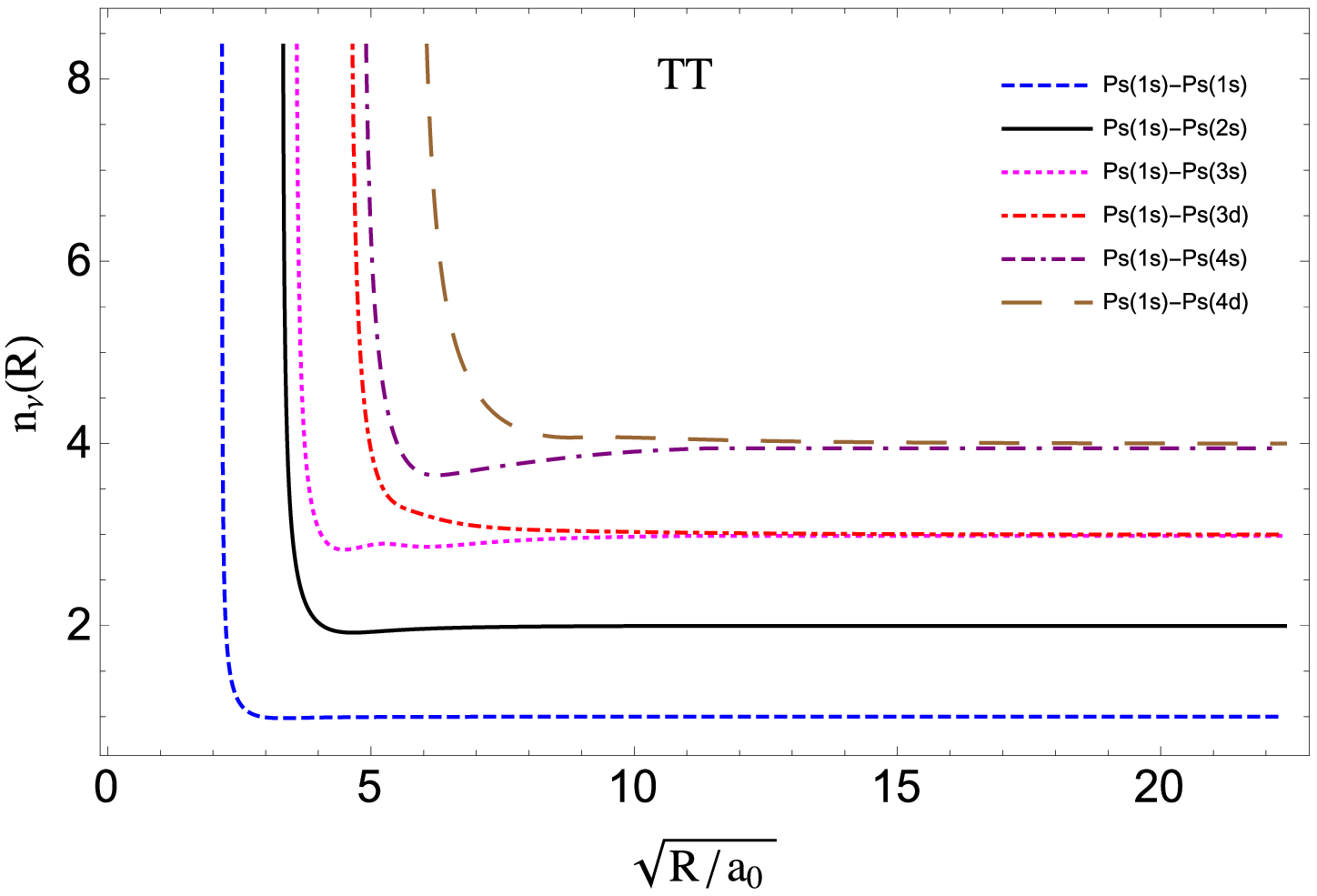}}

    \subfigure[]{\includegraphics[width=1\columnwidth]{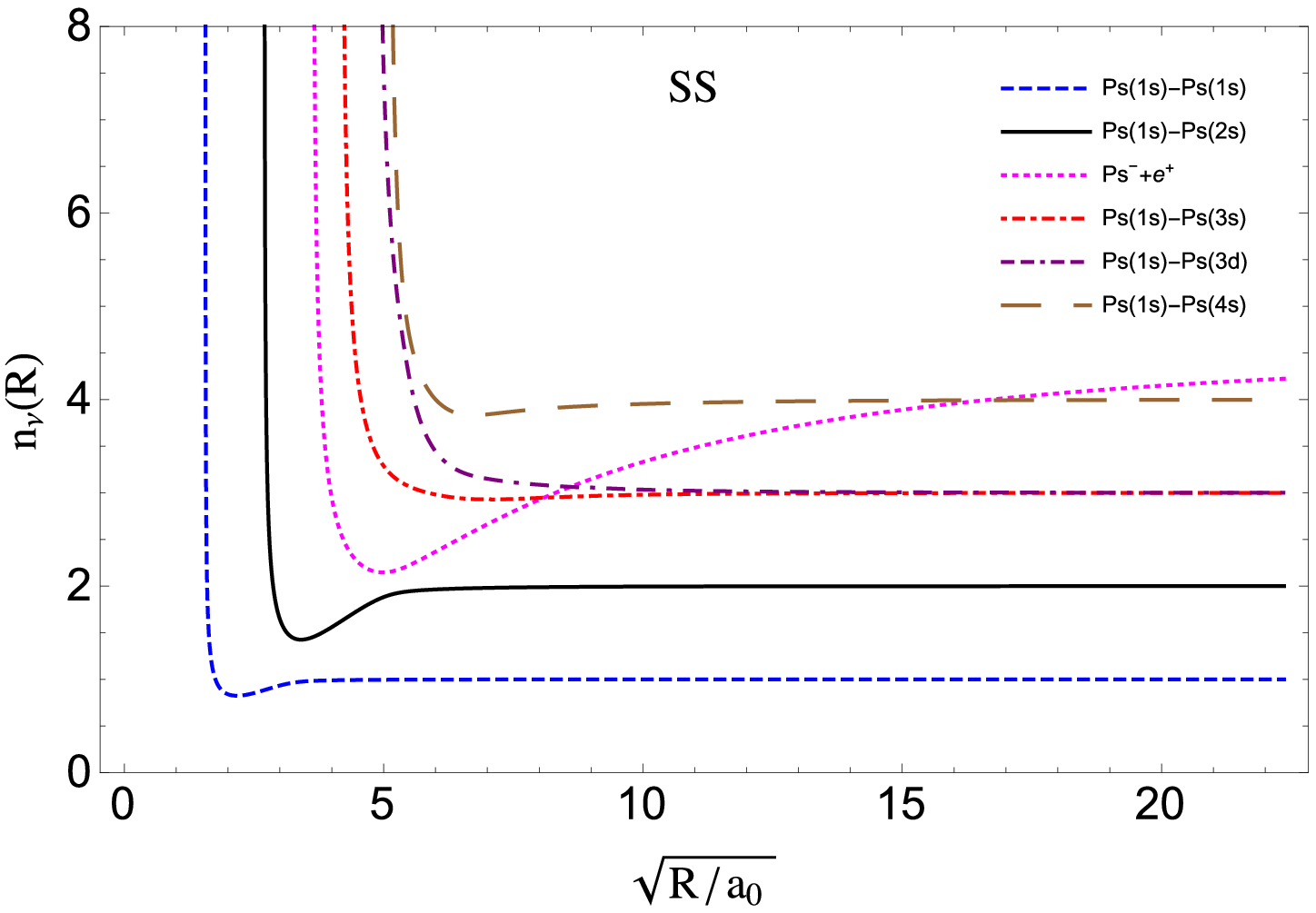}}
    
    \subfigure[]{\includegraphics[width=1\columnwidth]{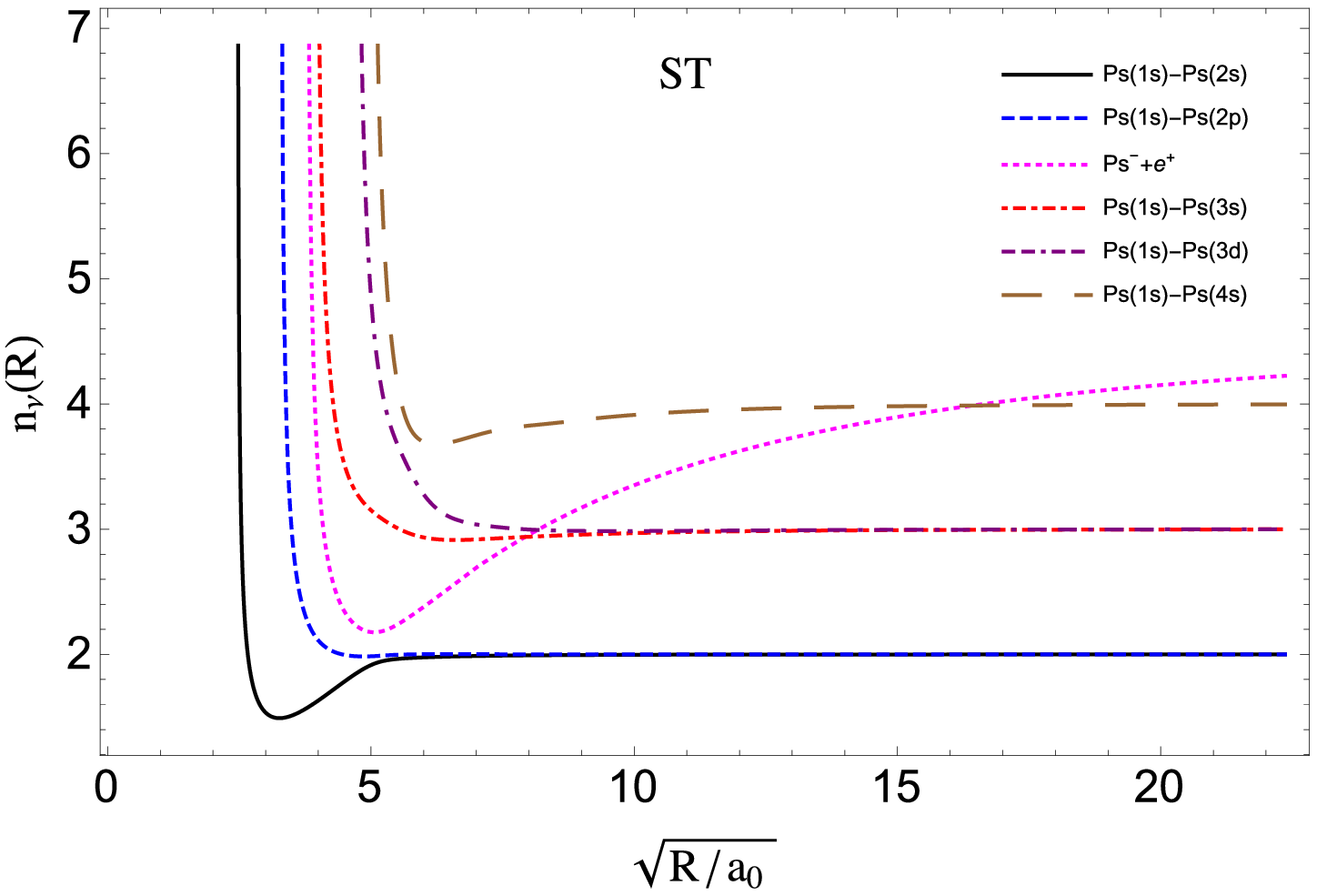}}
    
    \caption{The first 6 adiabatic potential curves for the TT (a), SS (b), and ST (c) spin configurations. These are plotted such that they approach an effective principal quantum number of one Ps dimer and the ground state of the other Ps dimer (see Eq. $14$ in \cite{e_+e_-Daily}). The solid curve represents the 1s2s channel of interest in all plots.}
    \label{fig:adiabaticpotentials}
\end{figure}

The adiabatic potentials $U_{\nu}(R)$ were calculated by expanding the basis functions $\phi_{\nu}(R,\Omega)$ in the explicitly correlated Gaussian basis then diagonalizing the adiabatic Hamiltionian (see \cite{e_+e_-Daily,DailyGreene2014,GreeneStecher2009,RakshitBlume,Mitroy2013}). The potential curves used in the scattering calculations within this paper are for the parity eigenvalues of +1 and total angular momentum L=0. Thus, from Eq. 19 of \cite{e_+e_-Daily}, the potential curves calculated are for states that represent, in the asymptotic limit, those containing a Ps atom in the ground state and a Ps atom in an excited state. The potentials for the SS and TT spin configurations are computed for eigenstates of charge conjugation with eigenvalue +1, projecting out states of Ps in odd $l$-orbitals. Figure \ref{fig:adiabaticpotentials} shows the first 6 calculated adiabatic potentials for the TT (a), SS (b), and ST (c) spin representations.

From Fig. \ref{fig:adiabaticpotentials} (a), the dashed, solid, dotted, and dash-dash-dotted lines correspond to principal quantum numbers of $n=1,2,3$ and $4$ in the asymptotic limit with the dashed and solid lines corresponding to the asymptotic 1$s$1$s$ and 1$s$2$s$ channels, respectively. In Fig. \ref{fig:adiabaticpotentials} (b), the dashed, solid, dash-dotted, and medium-dashed lines correspond to principal quantum numbers of $n=1,2,3$ and $4$ in the asymptotic limit with the dashed and solid lines corresponding to the asymptotic 1s1s and 1s2s channels, respectively. The dotted curve in Fig. \ref{fig:adiabaticpotentials} (b) corresponds to ion-pair potentials when diabatically connected through avoided crossings. Similarly, in Fig. \ref{fig:adiabaticpotentials} (c), the solid, dashed, dash-dotted, and dash-dash-dotted and medium-dashed lines correspond to the $1s2s$, $1s2p$, $1s3s$, $1s3d$ and $1s4s$ asymptotic channels, respectively while the dotted curve represents the ion-pair breakup channel. Figure \ref{fig:adiabaticpotentials} (a) does not contain an ion-pair channel because $Ps^+$ and $Ps^-$ are unstable when the two identical particles in the system are in the triplet spin state.

The potentials in Fig. \ref{fig:adiabaticpotentials} were calculated by diagonalizing the adiabatic Hamiltonian using 247, 300, and 280 basis functions for SS, TT, and ST spin configurations, respectively. The potentials were calculated out to a hyperradius of $R=500a_0$, where $a_0$ is defined as 1 Bohr in atomic units (or $5.29177\times10^{-11}$ m in S.I. units). Due to the basis sizes used, the results for the potentials and the non-adiabatic coupling matrices were not well converged at large-$R$ values ($R\geq100a_0$). Thus, power-law fits were performed and spliced to the expected behavior (see \cite{DailyAsym}) to ensure convergence.

The SS and ST potentials show features of avoided crossings which trace out ionic channels, representing the asymptotic break-up channel with energy $E_{ionic}=-0.262E_h$. Near the avoided crossings, it is observed that the non-adiabatic $P_{\mu\nu}(R)$ matrix elements are sharply peaked, enabling a Landau-Zener analysis of the adiabatic and diabatic transition probabilities. The $P_{\mu\nu}(R)$ matrix elements near the avoided crossings follow a typical Lorentzian behavior, therefore Landau-Zener parameters are extracted from a fit to Eq. $12$ and non-adiabatic transition probabilities are calculated from Eq. 7 of \cite{ClarkLZ}. The energy gap at the avoided crossings are $4.2481\times10^{-4}$ and $6.6613\times10^{-4}$ Hartrees for the SS and ST potentials, respectively. The non-adiabatic transition probabilities at the avoided crossings indicate that diabatic transitions are likely to occur. Thus, the potentials are connected diabatically through the close avoided crossings, as indicated by the curve-crossings in Figs. \ref{fig:adiabaticpotentials} (b) and (c).

\section{Numerical Analysis: Low-Energy Multichannel Scattering}
\label{sec:numanalysis}

\begin{figure}[htb]
\centering
\subfigure[]{\includegraphics[width=1\columnwidth]{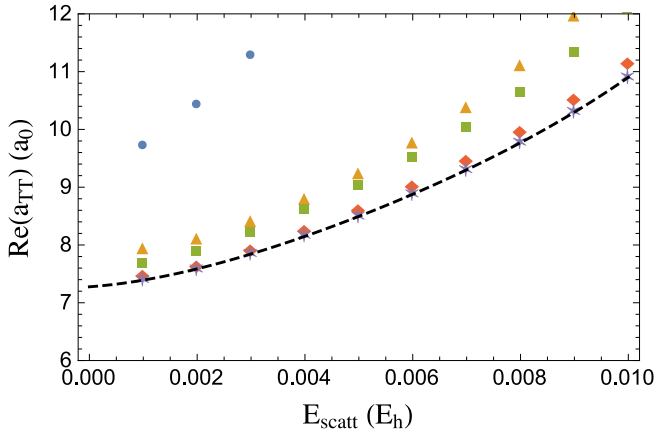}}

\subfigure[]{\includegraphics[width=1\columnwidth]{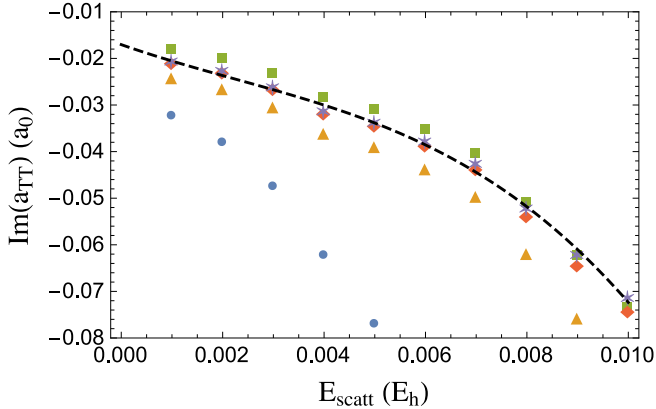}}
\caption{The TT s-wave scattering lengths are plotted versus the collision energy above the $1s-2s$ threshold in atomic units. Panel (a) shows the real part and panel (b) shows the imaginary part of the scattering length. Circles, triangles, squares, diamonds, and asterisks represents the inclusion of 2--6 channels, respectively. A $5^\mathrm{th}$-order polynomial fit to the $6$-channel calculation is shown as the dashed line.}
\label{fig:scattlengthTT}
\end{figure}

In this section, the low-energy scattering calculations for scattering in the 1s2s dimer-dimer breakup channel (dimer-dimer threshold energy of $E_{1s2s}=-0.3125E_h$) are described. The scattering lengths for the TT, SS and ST spin configurations are calculated, and an appropriate recoupling is carried out, in order to describe scattering of the physical spin states. Figure \ref{fig:scattlengthTT} shows the real and imaginary parts of the scattering length for the triplet-triplet spin symmetry versus scattering energy in the $1s-2s$ breakup channel.

\begin{figure}[!ht]
\centering
\subfigure[]{\includegraphics[width=1\columnwidth]{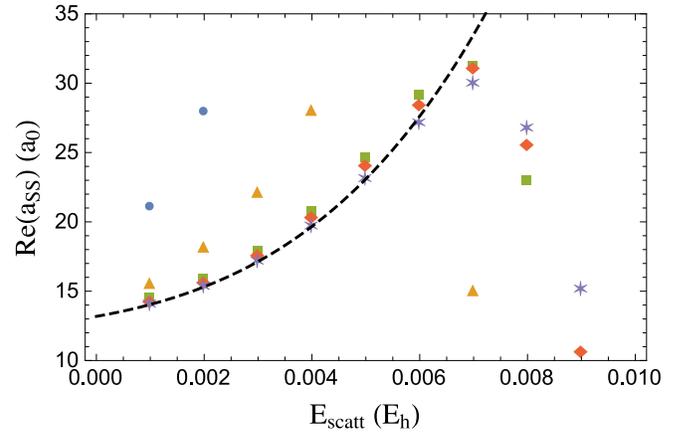}}

\subfigure[]{\includegraphics[width=1\columnwidth]{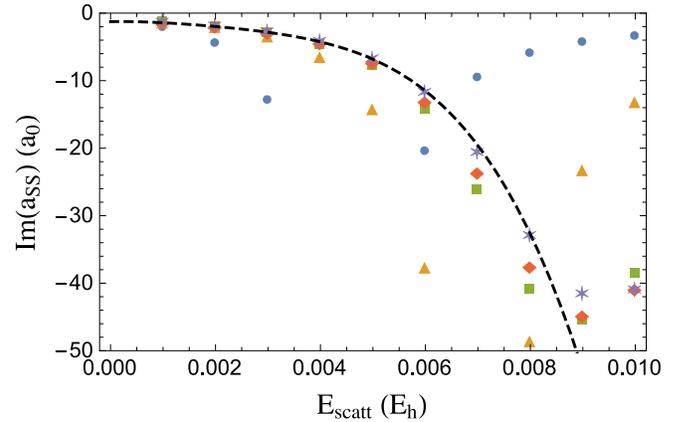}}
\caption{The SS s-wave scattering lengths are plotted versus the collision energy above the 1s2s threshold in atomic units. Panel (a) shows the real part and (b) shows the imaginary part of the scattering length. Circles, triangles, squares, diamonds, and asterisks represents the inclusion of 2--6 channels, respectively. A $5^\mathrm{th}$-order polynomial fit to the $6$-channel calculation is shown as the dashed line.}
\label{fig:scattlengthSS}
\end{figure}

In Fig.~\ref{fig:scattlengthTT}, the data corresponding to circles, triangles, squares, diamonds, and stars represents the inclusion of 2, 3, 4, 5, and 6 channels, respectively. The scattering lengths are calculated by propagating the R matrix (see ref. \cite{Wang} for further details) out to a matching radius of $R=$~$500a_0$, then extrapolating out to an infinite matching radius. Each curve represents the inclusion of 2--6 channels with the first two corresponding to the open 1s1s and 1s2s channels, with the rest being closed channels. The behavior of these curves indicates that the scattering length is converging to a constant value at each scattering energy with the inclusion of more closed channels. The scattering length is also obtained for the SS spin configuration. The numerical results for both the real and imaginary scattering lengths are shown in Fig. \ref{fig:scattlengthSS}.

\begin{figure}[!ht]
\centering
\subfigure[]{\includegraphics[width=1\columnwidth]{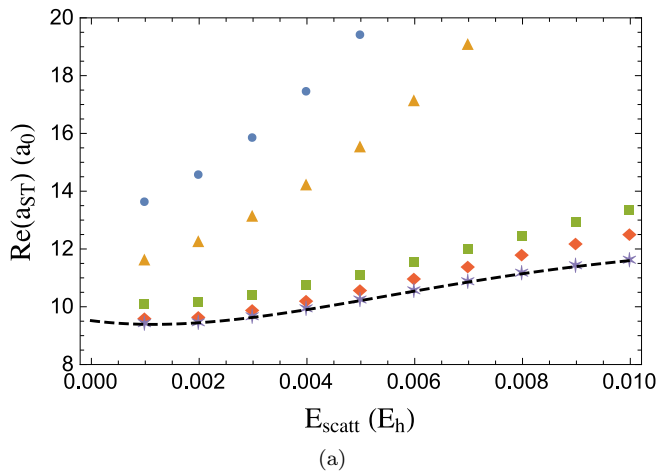}}

\subfigure[]{\includegraphics[width=1\columnwidth]{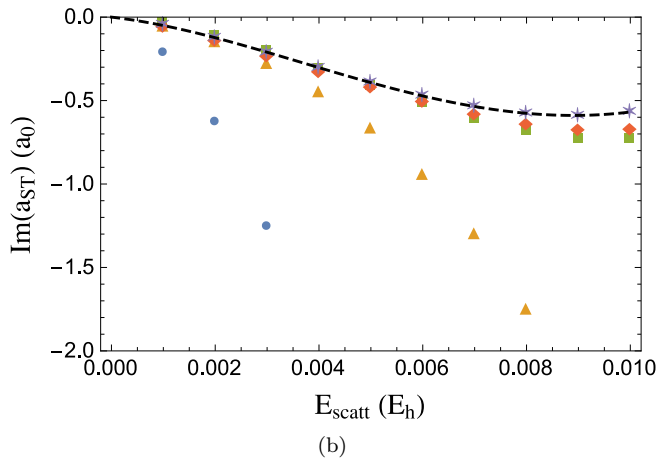}}

\caption{The real (a) and imaginary (b) ST s-wave scattering lengths plotted versus scattering energy above the 1s2s threshold in atomic units. Circles, triangles, squares, diamonds, and asterisks represents the inclusion of 2--6 channels, respectively. A $5^\mathrm{th}$-order polynomial fit to the $6$-channel calculation is shown as the dashed line.}
\label{fig:scattlengthST}
\end{figure}

In Fig.~\ref{fig:scattlengthSS}, the data corresponding to circles, triangles, squares, diamonds, and stars represents the inclusion of 2, 3, 4, 5, and 6 channels, respectively. It can be seen that the scattering length appears to converge reasonably well when more channels are included in the calculation. The values displayed are from extrapolation out to an infinite matching radius as in the TT case. The ST scattering length is extrapolated the same way as the previous cases and is shown in Fig. \ref{fig:scattlengthST}. 

\begin{figure}[!hb]
    \centering
    \subfigure[]{\includegraphics[width=1\columnwidth]{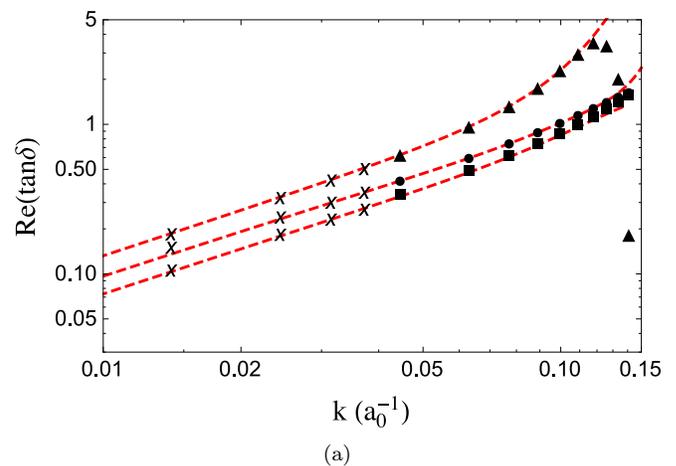}}
    
    \subfigure[]{\includegraphics[width=1\columnwidth]{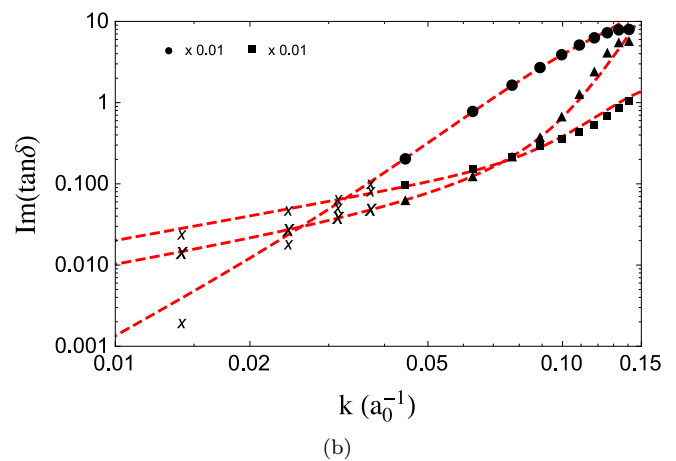}}
    \caption{The real (a) and imaginary (b) parts of $\tan{(\delta)}$ with respect to wavenumber in the 1s2s channel plotted on a log-log scale. Circles, triangles, and squares represent the ST, SS, and TT spin configurations, respectively. The dashed lines are fits to the Wigner threshold law. Crosses are numerical results computed at smaller energies not used in the fitting procedure. The legend in (b) gives a multiplication factor for the specified data.}
    \label{fig:tanscattcombined}
\end{figure}

For the ST spin configuration, the zero-energy scattering length is real due to short-range coupling between the 1$s$2$s$ and 1$s$2$p$ channels, which is evident in Fig. \ref{fig:scattlengthST} (b). From Figs. \ref{fig:scattlengthTT}, \ref{fig:scattlengthSS}, and \ref{fig:scattlengthST}, the curve corresponding to the inclusion of 6 channels appears to be converged, and is thus used to calculate the $s$-wave scattering length for all spin configurations. A $5^{\mathrm{th}}$-order polynomial fit to the data for 6 included channels is shown and used as a first pass at obtaining the zero-energy scattering lengths for TT, SS, and ST spin configurations. As another method in determining the $s$-wave scattering lengths, the $k$-dependence of $\tan{(\delta)}$ is studied, where $k$ is the wavenumber and $\delta$ is the phase-shift in the 1s2s breakup channel. Figure \ref{fig:tanscattcombined} shows this $k$-dependence for both the real and imaginary parts of $\tan{(\delta)}$ for the SS, TT, and ST spin configurations.

From Fig \ref{fig:tanscattcombined}, it can be seen that $\tan{(\delta)}$ approaches zero as a power of $k$ with a slope representing the order of the power on a log scale, which is the expected behavior in accordance to the Wigner threshold law. Thus, a simple power-law least-squares fit is performed for each of the plots to extract the scattering length. The first 4 data points were fitted to a function of the form:
\begin{equation}
    \tan{(\delta)}=ak+bk^3+ck^5+dk^7+O(k^9)
    \label{eq:wignerfit}
\end{equation}
where the scattering length is given by the parameter $a$. Fitting the data shown in Fig. \ref{fig:tanscattcombined} to Eq. \ref{eq:wignerfit}, in combination with the previous method, results in $a_{\mathrm{TT}}=$~$7.3(2)a_0-i0.02(1)a_0$, $a_{\mathrm{SS}}=$~$13.2(2)a_0-i0.9(2)a_0$, and $a_{\mathrm{ST}}=$~$9.7(2)a_0$ for the TT, SS, and ST $s$-wave scattering lengths, respectively. Further evidence of zero imaginary-part of the scattering length in the ST spin data (circles) is clearly represented in Fig. \ref{fig:tanscattcombined} (b). The data converges to a line of slope 3 on the log-log scale, which is the lowest power of $k$ in the Wigner threshold law behavior, leading to zero imaginary scattering length.

\begin{figure}
    \centering
    \includegraphics[width=1\columnwidth]{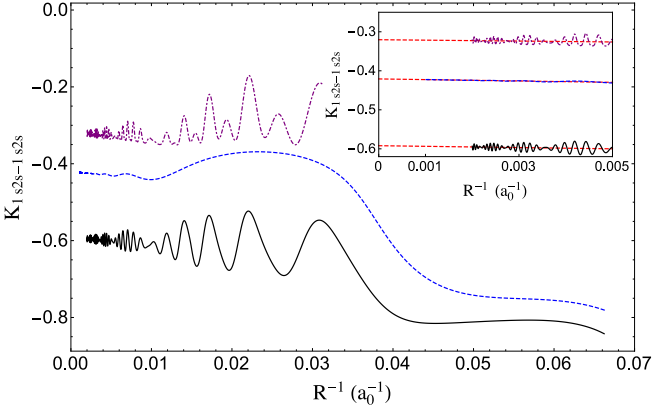}
    \caption{The $\underline{K}$-matrix elements for the TT (dot-dashed), SS (solid), and ST (dashed) spin configurations in the 1$s$--2$s$ elastic channel at $E_{scatt}$=0.001$E_h$ with the inclusion of 6 channels. The inset shows the linear fit used to extract the large-$R$ values.}
    \label{fig:Kmatrixcombined}
\end{figure}

The error estimates for the recorded scattering lengths are derived from factors such as oscillations in the $\underline{K}$-matrix due to non-adiabatic coupling between channels, and uncertainties in the least-square fitting of the data for $\tan(\delta)$ to the Wigner threshold law and the 6-channel, energy-dependent scattering length data.
When plotted as a function of $R^{-1}$, the $\underline{K}$-matrix elements oscillate with increasing frequency as $R\rightarrow\infty$, as is shown in Fig. \ref{fig:Kmatrixcombined} for $\underline{K}$-matrix elements in the 1$s$--2$s$ elastic channel. Also, a linear fit is shown for the inclusion of 6 channels at $E_{scatt}=0.001E_H$ for the SS (solid), TT (dot-dashed), and ST (dashed) spin configurations. The other elements of the $\underline{K}$-matrix are treated similarly, but are not shown. The oscillatory behavior in the $\underline{K}$-matrix is a result of coupling between channels through the $\underline{P}$ and $\underline{Q}$ matrices. When the coupling is artificially turned off, it is observed that the oscillations are removed, as was also observed in \cite{e_+e_-Daily}. The quoted error derives from an analysis of a linear fit to this data to extract the large-$R$ matrix elements extrapolated to an infinite matching radius. The fit is performed over a range $200a_0\leq R \leq500a_0$ to average over many oscillations.

\section{Spin Recoupling}
\label{sec:recoupling}
As previously stated, the scattering lengths discussed to this point are for the total spin configurations of the coupled electrons and coupled positrons, i.e. with good quantum numbers $S_+,S_-$ which are constants of the motion for the non-relativistic Hamiltonian that neglects annihilation and spin-dependent interactions. The physical spin states are those of the two positronium atoms. Re-coupling the spin states of the two positronium atoms to the electron-electron and positron-positron spin representation will yield expressions for the experimentally measurable scattering lengths in terms of the values calculated in this paper. 

The recoupling transformation of the spin states can be expressed through a unitary transformation. The unitary transformation matrix elements between these two representations is constructed from the overlap of the corresponding symmetrized channel functions. In the asymptotic limit ($R\rightarrow\infty$) the symmetrized channel functions in both representations are given by Eqs. \ref{eq:symchfun} and \ref{eq:symchfun1}.
\begin{widetext}
\begin{multline}
    \hat{S}\left|\Phi_{\nu}(R,\Omega)\right\rangle\rightarrow\frac{1}{\sqrt{4(1+(-1)^{l_{3}+S_{+}+S_{-}}\delta_{n_1,n_2}\delta_{l_1,l_2}\delta_{m_1,m_2})}}[\left|n_{1}l_{1}(13)\right\rangle\left|n_{2}l_{2}(24)\right\rangle\left|\hat{\rho}_{13}\hat{\rho}_{24}\hat{\rho}_{3}\right\rangle\\+(-1)^{l_{3}+S_{+}+S_{-}}\left|n_{1}l_{1}(24)\right\rangle\left|n_{2}l_{2}(13)\right\rangle\left|\hat{\rho}_{24}\hat{\rho}_{13}\hat{\rho}_{3}\right\rangle\\+(-1)^{S_{+}}\left|n_{1}l_{1}(23)\right\rangle\left|n_{2}l_{2}(14)\right\rangle\left|\hat{\rho}_{23}\hat{\rho}_{14}\hat{\rho}_{\bar{3}}\right\rangle\\+(-1)^{l_{3}+S_{-}}\left|n_{1}l_{1}(14)\right\rangle\left|n_{2}l_{2}(23)\right\rangle\left|\hat{\rho}_{14}\hat{\rho}_{23}\hat{\rho}_{\bar{3}}\right\rangle]\\\times\left|(e_1^{+}e_2^{+})S_{+}(e_3^{-}e_4^{-})S_{-};S_{\mathrm{tot}}\right\rangle
    \label{eq:symchfun}
\end{multline}
\begin{multline}
    \hat{S}\left|\Psi_{\nu}(R,\Omega)\right\rangle\rightarrow\\\frac{1}{\sqrt{4(1+(-1)^{l_{3}+S_{\mathrm{tot}}}\delta_{S_{13},S_{24}}\delta_{n_1,n_2}\delta_{l_1,l_2}\delta_{m_1,m_2})}}[\left|n_{1}l_{1}(13)\right\rangle\left|n_{2}l_{2}(24)\right\rangle\left|\hat{\rho}_{13}\hat{\rho}_{24}\hat{\rho}_{3}\right\rangle\left|(e_1^{+}e_3^{-})S_{13}(e_2^{+}e_4^{-})S_{24};S_{\mathrm{tot}}\right\rangle\\+(-1)^{l_{3}}\left|n_{1}l_{1}(24)\right\rangle\left|n_{2}l_{2}(13)\right\rangle\left|\hat{\rho}_{24}\hat{\rho}_{13}\hat{\rho}_{3}\right\rangle\left|(e_2^{+}e_4^{-})S_{13}(e_1^{+}e_3^{-})S_{24};S_{\mathrm{tot}}\right\rangle\\-\left|n_{1}l_{1}(23)\right\rangle\left|n_{2}l_{2}(14)\right\rangle\left|\hat{\rho}_{23}\hat{\rho}_{14}\hat{\rho}_{\bar{3}}\right\rangle\left|(e_2^{+}e_3^{-})S_{13}(e_1^{+}e_4^{-})S_{24};S_{\mathrm{tot}}\right\rangle\\-(-1)^{l_{3}}\left|n_{1}l_{1}(14)\right\rangle\left|n_{2}l_{2}(23)\right\rangle\left|\hat{\rho}_{14}\hat{\rho}_{23}\hat{\rho}_{\bar{3}}\right\rangle\left|(e_1^{+}e_4^{-})S_{13}(e_2^{+}e_3^{-})S_{24};S_{\mathrm{tot}}\right\rangle]
    \label{eq:symchfun1}
\end{multline}
\end{widetext}
where the $n$'s and $l$'s represent the principal and angular momentum quantum numbers for the two bound positronium atoms, $\left|\hat{\rho}_{13}\hat{\rho}_{24}\hat{\rho}_{3}\right\rangle$ represents a coupled spherical harmonic (see \cite{DailyAsym,DailyGreene2014}), and $S_{13}$ and $S_{24}$ are the spins of the positronium atoms. The states given by Eqs. \ref{eq:symchfun} and \ref{eq:symchfun1} are zero when the Kronecker-delta factors are $1$ and the phase factors are $-1$. The physical spin states are eigenfunctions of charge conjugation with eigenvalue $(-1)^{S_{13}+l_1+S_{24}+l_2}$. The symmetrized overlap matrix element for positronium atoms in arbitrary orbitals is given by Eq. \ref{eq:transformation}:

\begin{widetext}
\begin{multline}
    \left\langle\hat{S}\Phi_{\nu}(R,\Omega;(S_{+}S_{-})S_{\mathrm{tot}})\right|\left.\hat{S}\Psi_{\nu}(R,\Omega;(S_{13}S_{24})S_{\mathrm{tot}})\right\rangle=\\\sqrt{1+(-1)^{l_3+S_{+}+S_{-}}\delta_{n_1,n_2}\delta_{l_1,l_2}\delta_{m_1,m_2}}\left(1-\left(1-\sqrt{\frac{1}{2}}\right)\left(\frac{1+(-1)^{l_3+S_{\mathrm{tot}}}}{2}\right)\delta_{S_{13},S_{24}}\delta_{n_1,n_2}\delta_{l_1,l_2}\delta_{m_1,m_2}\right)\times\\\times\sqrt{(2S_++1)(2S_-+1)(2S_{13}+1)(2S_{34}+1)}
{\left\{ {\begin{array}{ccc}
   1/2 & 1/2 & S_{\mathrm{+}}\\
   1/2 & 1/2 & S_{\mathrm{-}}\\
   S_{13} & S_{24} & S_{tot}\\
  \end{array} } \right\}}
  \label{eq:transformation}
\end{multline}
\end{widetext}
where the $\left\{\right\}$ symbol represents a Wigner 9j symbol \cite{AngMomentum,burke}. 

For some experiments it would be useful to know the scattering properties in the uncoupled positronium spin basis, $\left|Ps(n_1l_1m_1)[S_{13},M_{S_{13}}]Ps(n_2l_2m_2)[S_{24},M_{S_{24}}]\right\rangle$. The properly symmeterized matrix elements in this representation are modified slightly from those in Eq. \ref{eq:transformation}. The matrix elements for the uncoupled spin representation are shown in Eq. \ref{eq:uncoupledtransform}.
\begin{widetext}
\begin{multline}
    \left\langle\hat{S}\Phi_{\nu}(R,\Omega;(S_{+}S_{-})S_{\mathrm{tot}})\right|\left.\hat{S}\Psi_{\nu}(R,\Omega;(S_{13}M_{13}S_{24}M_{24})S_{\mathrm{tot}}M_{\mathrm{tot}})\right\rangle=\\\sqrt{1+(-1)^{l_3+S_{+}+S_{-}}\delta_{n_1,n_2}\delta_{l_1,l_2}\delta_{m_1,m_2}}\left(1-\left(1-\sqrt{\frac{1}{2}}\right)\left(\frac{1+(-1)^{l_3}}{2}\right)\delta_{S_{13},S_{24}}\delta_{M_{13},M_{24}}\delta_{n_1,n_2}\delta_{l_1,l_2}\delta_{m_1,m_2}\right)\\\times C_{S_{13}M_{13},S_{24}M_{24}}^{S_{\mathrm{tot}}M_{\mathrm{tot}}}\sqrt{(2S_++1)(2S_-+1)(2S_{13}+1)(2S_{34}+1)}
{\left\{ {\begin{array}{ccc}
   1/2 & 1/2 & S_{\mathrm{+}}\\
   1/2 & 1/2 & S_{\mathrm{-}}\\
   S_{13} & S_{24} & S_{tot}\\
  \end{array} } \right\}}
  \label{eq:uncoupledtransform}
\end{multline}
\end{widetext}
In Eq. \ref{eq:uncoupledtransform}, $C_{S_{13}M_{13},S_{24}M_{24}}^{S_{\mathrm{tot}}M_{\mathrm{tot}}}$ represents a Clebsch-Gordan coefficient.

With the transformation matrix element expressed in Eq. \ref{eq:transformation}, the relevant matrices describing the scattering event in the positronium spin basis can easily be formed through a unitary transformation constructed from these elements. The general transformation of matrix elements from one basis to another is described in Eq. \ref{eq:mateltrans}:
\begin{widetext}
\begin{multline}
    \left\langle\hat{S}\Psi_{\nu'}((S'_{13}S'_{24})S_{\mathrm{tot}})\right|\hat{A}\left|\hat{S}\Psi_{\nu}((S_{13}S_{24})S_{\mathrm{tot}})\right\rangle=\\\sum_{S_{+},S_{-}}\left\langle\hat{S}\Psi_{\nu'}((S'_{13}S'_{24})S_{\mathrm{tot}})\right|\left.\hat{S}\Phi_{\nu'}((S_{+}S_{-})S_{\mathrm{tot}})\right\rangle A_{v,v'}^{S_{+},S_{-}} \left<\hat{S}\Phi_{\nu}((S_{+}S_{-})S_{\mathrm{tot}})\right|\left.\hat{S}\Psi_{\nu}((S_{13}S_{24})S_{\mathrm{tot}})\right\rangle
\label{eq:mateltrans}
\end{multline}
\end{widetext}
where $A_{\nu,\nu'}^{S_{+},S_{-}}$ is the $\nu,\nu'$ matrix element of matrix $\underbar{A}$ in the $S_{+}$ and $S_{-}$ spin basis. The space variables are suppressed in Eq. \ref{eq:mateltrans}. The unitary transformation of matrix elements to the uncoupled Ps spin basis is given by inserting Eq. \ref{eq:uncoupledtransform} into Eq. \ref{eq:mateltrans} and summing over $S_{\mathrm{tot}}$ with $M_{S_{\mathrm{tot}}}=M_{S_{13}}+M_{S_{24}}$. The matrix elements of interest are those of the scattering matrix $\underbar{S}$, the transition matrix $\underbar{T}$, and the scattering length matrix $\underbar{a}$. The results for Ps(1s)-Ps(2s) T-Matrix elements in different uncoupled spin states are shown in Table \ref{table:tmatrix}.
\begin{table}[ht]
\caption{T-matrix element transformations for uncoupled Ps spin states. Processes not shown are zero.}
\centering
\label{table:tmatrix}
\resizebox{1\columnwidth}{!}{\begin{tabular}{l|l|lll}
    $\#$ & $S'_{13}M'_{13},S'_{24}M'_{24} \leftrightarrow S_{13}M_{13},S_{24}M_{24}$ & $T_{1s1s\leftrightarrow1s1s}$ & $T_{1s2s\leftrightarrow1s1s}$ & $T_{1s2s\leftrightarrow1s2s}$\\
    \hline
    $1$ & $(1,\pm1),(1,\pm1)\leftrightarrow(1,\pm1),(1,\pm1)$  & $T_{TT}$ & $T_{TT}$ & $T_{TT}$\\ 
    $2$ & $(1,\pm1),(1,\mp1)\leftrightarrow(1,\pm1),(1,\mp1)$  & $\frac{1}{2}(T_{SS}+T_{TT})$ & $\frac{1}{2\sqrt{2}}(T_{SS}+T_{TT})$ & $\frac{1}{4}(T_{SS}+T_{TT})+\frac{1}{2}T_{ST}$ \\
    $3$ & $(1,\pm1),(1,\mp1)\leftrightarrow(1,\mp1),(1,\pm1)$  & $\frac{1}{2}(T_{SS}+T_{TT})$ & $\frac{1}{2\sqrt{2}}(T_{SS}+T_{TT})$ & $\frac{1}{4}(T_{SS}+T_{TT})-\frac{1}{2}T_{ST}$ \\
    $4$ & $(1,\mp1),(1,\pm1)\leftrightarrow(1,0),(1,0)$ & $\frac{1}{2\sqrt{2}}(T_{TT}-T_{SS})$ & $\frac{1}{4}(T_{TT}-T_{SS})$ & $\frac{1}{4}(T_{TT}-T_{SS})$\\
    $5$ & $(1,\mp1),(1,\pm1)\leftrightarrow(0,0),(0,0)$  & $\frac{1}{2\sqrt{2}}(T_{SS}-T_{TT})$ & $\frac{1}{4}(T_{SS}-T_{TT})$ & $\frac{1}{4}(T_{SS}-T_{TT})$ \\
    $6$ & $(1,\pm1),(1,0)\leftrightarrow(1,\pm1),(1,0)$  & $T_{TT}$ & $\frac{1}{\sqrt{2}}T_{TT}$ & $\frac{1}{2}(T_{TT}+T_{ST})$ \\
    $7$ & $(1,\pm1),(1,0)\leftrightarrow(1,0),(1, \pm1)$  & $T_{TT}$ & $\frac{1}{\sqrt{2}}T_{TT}$ & $\frac{1}{2}(T_{TT}-T_{ST})$ \\
    $8$ & $(1,\pm1),(0,0)\leftrightarrow(1,\pm1),(0,0)$  & $T_{TT}$ & $\frac{1}{\sqrt{2}}T_{TT}$ & $\frac{1}{2}(T_{TT}+T_{ST})$ \\
    $9$ & $(1,\pm1),(0,0)\leftrightarrow(0,0),(1,\pm1)$  & $T_{TT}$ & $\frac{1}{\sqrt{2}}T_{TT}$ & $\frac{1}{2}(T_{TT}-T_{ST})$  \\
    $10$ & $(1,0),(1,0)\leftrightarrow(1,0),(1,0)$  & $\frac{1}{4}T_{SS}+\frac{3}{4}T_{TT}$ & $\frac{1}{4}T_{SS}+\frac{3}{4}T_{TT}$ & $\frac{1}{4}T_{SS}+\frac{3}{4}T_{TT}$ \\
    $11$ & $(1,0),(1,0)\leftrightarrow(0,0),(0,0)$  & $\frac{1}{4}(T_{TT}-T_{SS})$ & $\frac{1}{4}(T_{TT}-T_{SS})$ & $\frac{1}{4}(T_{TT}-T_{SS})$\\
    $12$ & $(1,0),(0,0)\leftrightarrow(1,0),(0,0)$  & $T_{TT}$ & $\frac{1}{\sqrt{2}}T_{TT}$ & $\frac{1}{2}(T_{TT}+T_{ST})$ \\
    $13$ & $(1,0),(0,0)\leftrightarrow(0,0),(1,0)$  & $T_{TT}$ & $\frac{1}{\sqrt{2}}T_{TT}$ & $\frac{1}{2}(T_{TT}-T_{ST})$ \\
    $14$ & $(0,0),(0,0)\leftrightarrow(0,0),(0,0)$  & $\frac{1}{4}T_{SS}+\frac{3}{4}T_{TT}$ & $\frac{1}{4}T_{SS}+\frac{3}{4}T_{TT}$ & $\frac{1}{4}T_{SS}+\frac{3}{4}T_{TT}$\\\hline
\end{tabular}}
\end{table}

Table \ref{table:tmatrix} displays the T-matrix elements for collisions of positronium atoms in different uncoupled spin states and excited states, specifically the 1$s$ and 2$s$ states, computed from Eq. \ref{eq:transformation}. Each column gives the positronium spin collision channel, and the energy state of each positronium in the collision, respectively. The T-matrix elements in the $1s1s\leftrightarrow1s1s$ collision channel matches those used in reference \cite{Ivanov}. It is observed that some spin collision channels are not allowed. This is, in part, due to antisymmetrization of the asymptotic channel functions in the positronium spin basis (see Eq. \ref{eq:symchfun1}), properties of the spin functions, and the restriction to $s$-wave collisions only, which has been imposed here.

Some of the spin processes are not allowed due to the conservation of the total spin projection quantum number, $M_{S_{tot}}$. Since $M_{S_{tot}}=M_{S_{13}}+M_{S_{24}}$, states of total spin and spin projection quantum numbers will only consist of uncoupled states where this condition is satisfied. Therefore, uncoupled spin states are orthogonal if $M_{S_{tot}}$ is not the same. Other processes are zero due to the symmetry requirements on the wavefunction and the Hamiltonian being invariant under charge conjugation. 

For $1s1s\leftrightarrow1s1s$ and $1s1s\leftrightarrow1s2s$ elastic and inelastic collisions, spin collisions are not allowed for processes where Ps spin states couple to states where $S_+\neq S_-$($S_{tot}=1$), due to symmetry requirements on the asymptotic $1s1s$ wavefunction. From Eq. \ref{eq:symchfun}, the wavefunction vanishes if $(-1)^{l_3+S_++S_-}=-1$. Since we are studying the system for total angular momentum, $L=0$, $l_3=0$ when coupled to $s$-orbitals. Therefore, the wavefunction vanishes for odd values of $S_++S_-$, meaning either $S_+=1$ and $S_-=0$ or vice versa. 

The $1s2s\leftrightarrow1s2s$ elastic channel has vanishing transformed T-matrix elements for some spin collision processes for reasons described previously and others because the Hamiltonian is invariant under charge conjugation. Collisions with $M_{tot}=0$ and one Ps spin flips from the triplet to the singlet state or vice versa will not occur because this type of collision involves both symmetric and anti-symmetric spin states for $S_{tot}=1$. This will result in transformations proportional to the difference in ST and TS scattering properties in the $e^+$ $e^+$, $e^-$ $e^-$ spin coupled representation. Since the Hamiltonian is invariant under charge conjugation, the scattering properties of the ST and TS spin configurations are the same, leading to zero transformation elements.

In Table \ref{table:tmatrix}, some matrix element transformations are the same for different spin collisions. Some are trivially the same for symmetry arguments, while others are somewhat non-trivial(e.g. rows 10 and 14). The non-trivial cases result from neglecting Ps spin-spin interactions in the Hamiltonian, resulting in the scattering matrix elements being independent of the quantum number $S_{\mathrm{tot}}$. For completeness, the $1s2s\leftrightarrow1s2s$ collision channel, $s$-wave scattering lengths are tabulated in Table \ref{table:1s2sscattlengths} in the uncoupled Ps spin basis.
\begin{table}[ht]
\caption{Zero-energy 1$s$--2$s$ $s$-wave scattering lengths for uncoupled Ps spin states in atomic units. Processes not shown are zero.}
\centering
\label{table:1s2sscattlengths}
\resizebox{1\columnwidth}{!}{\begin{tabular}{l|l|l}
    $\#$ & $S'_{13}M'_{13},S'_{24}M'_{24} \leftrightarrow S_{13}M_{13},S_{24}M_{24}$ & $a_{1s-2s} (a_{0})$\\
    \hline
    $1$ & $(1,\pm1),(1,\pm1)\leftrightarrow(1,\pm1),(1,\pm1)$ & $7.3(2)-i0.02(1)$\\ 
    $2$ & $(1,\pm1),(1,\mp1)\leftrightarrow(1,\pm1),(1,\mp1)$ & $10.0(2)-i0.2(2)$ \\
    $6$ & $(1,\pm1),(1,0)\leftrightarrow(1,\pm1),(1,0)$  & $8.5(2)-i0.01(1)$ \\
    $8$ & $(1,\pm1),(0,0)\leftrightarrow(1,\pm1),(0,0)$ & $8.5(2)-i0.01(1)$ \\
    $10$ & $(1,0),(1,0)\leftrightarrow(1,0),(1,0)$ & $8.8(2)-i0.2(1)$ \\
    $12$ & $(1,0),(0,0)\leftrightarrow(1,0),(0,0)$ & $8.5(2)-i0.01(1)$ \\
    $14$ & $(0,0),(0,0)\leftrightarrow(0,0),(0,0)$ & $8.8(2)-i0.2(1)$\\\hline
\end{tabular}}
\end{table}

The results in Table \ref{table:1s2sscattlengths} can be used to give an estimate of the clock-shift frequency of a gas of 1$s$ and 2$s$ Ps. For a gas of spin-polarized triplet Ps, the clock-shift frequency per unit density is calculated to be $\frac{\Delta \nu_{1s-2s}}{n}=10.0(2)\times10^{-8} cm^{3}$-$Hz$, using the calculated TT 1$s$--2$s$ scattering length and the calculated 1$s$--1$s$ TT scattering length of $3.2a_{0}$ given in ref. \cite{e_+e_-Daily}.

\section{Cross-Sections}
\label{sec:crosssection}
In this section, partial cross-sections are calculated in the $1s2s\leftrightarrow1s2s$, $1s2s\rightarrow1s1s$ and $1s1s\leftrightarrow1s1s$ collision channels and different spin channels from the transformation of the T-matrix using Eq. \ref{eq:uncoupledtransform}. From the T-matrix elements, partial cross-sections can be determined in the standard way as $\sigma_{if}=\frac{\alpha}{k_{i}^2}\left|T_{if}\right|^2$. The factor of $\alpha$ is $4\pi$ for distinguishable $\rightarrow$ distinguishable and distinguishable $\rightarrow$ indistinguishable atom/particle collisions. For indistinguishable $\rightarrow$ indistinguishable and indistinguishable $\rightarrow$ distinguishable atom/particle collisions, $\alpha$ is $8\pi$ \cite{burke}.

\begin{figure}[!ht]
\centering
\subfigure[]{\includegraphics[width=1\columnwidth]{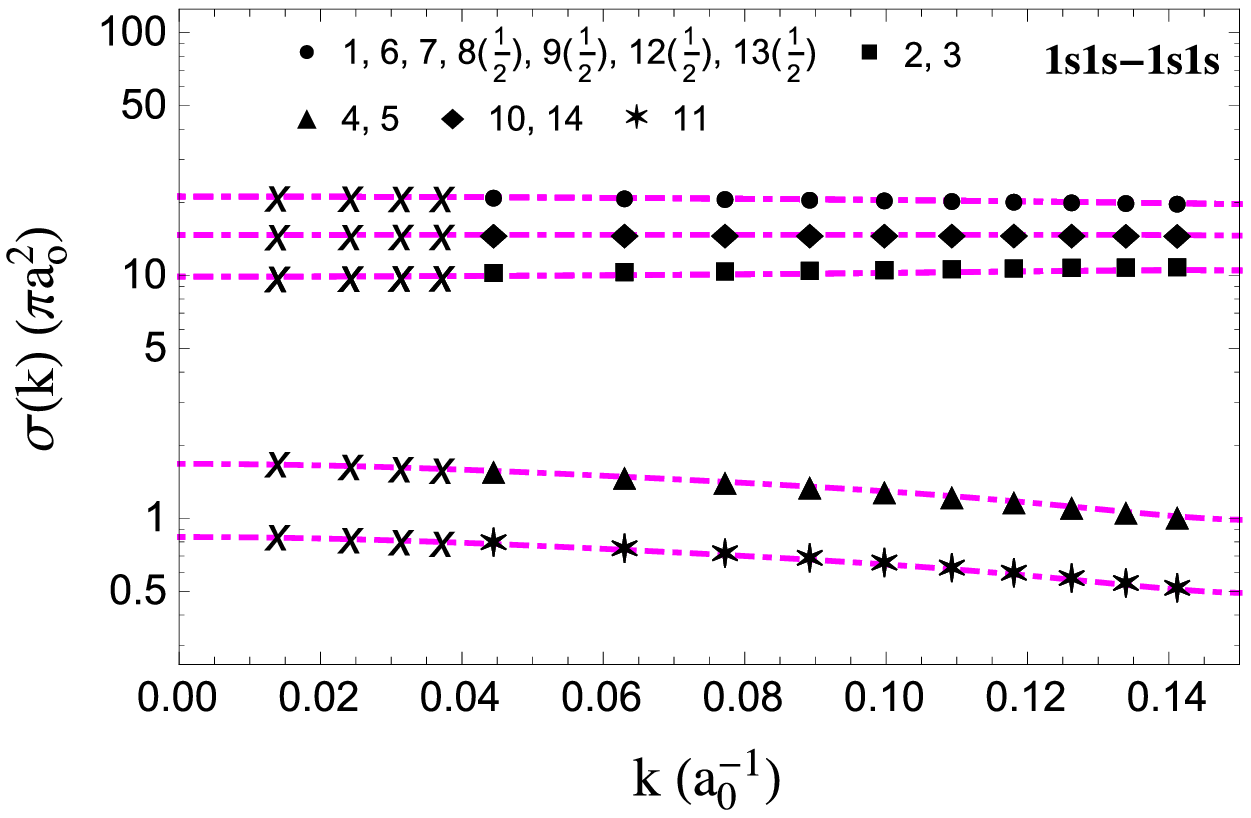}}

\subfigure[]{\includegraphics[width=1\columnwidth]{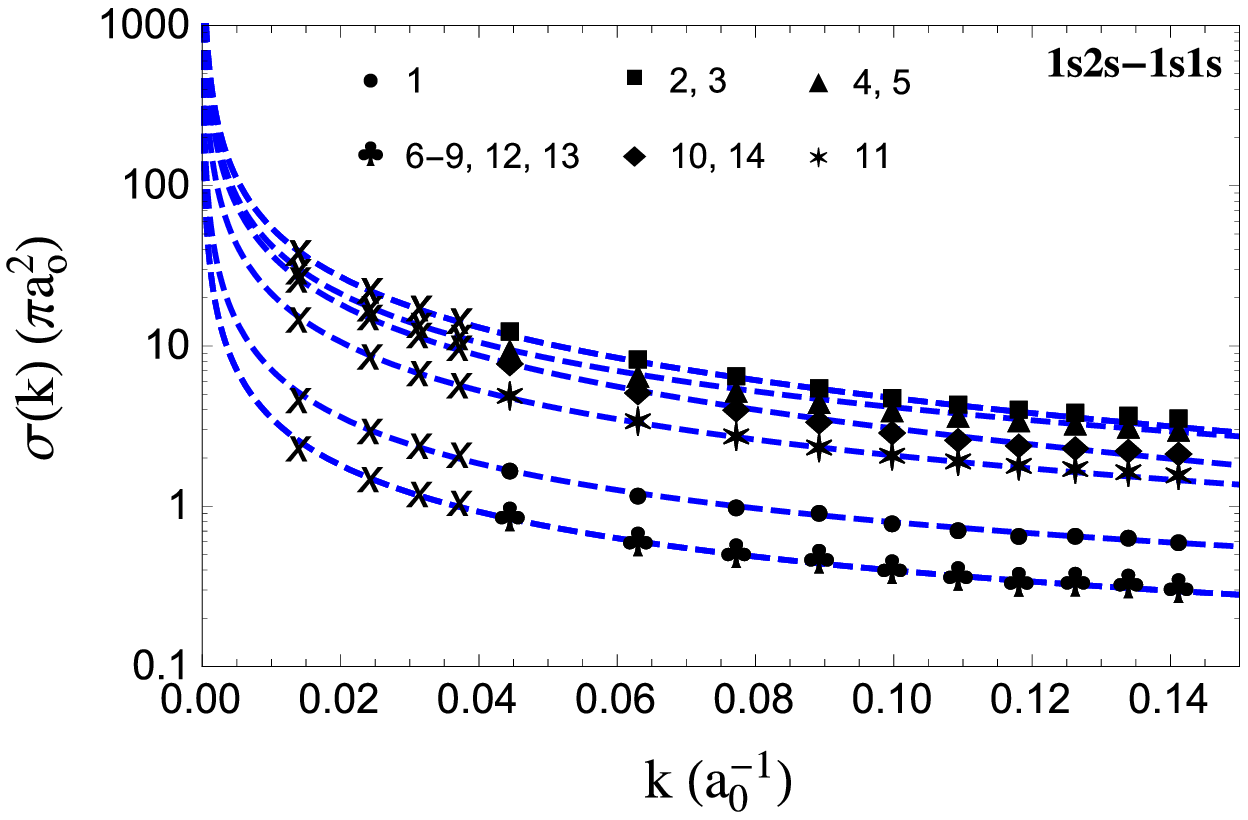}}

\subfigure[]{\includegraphics[width=1\columnwidth]{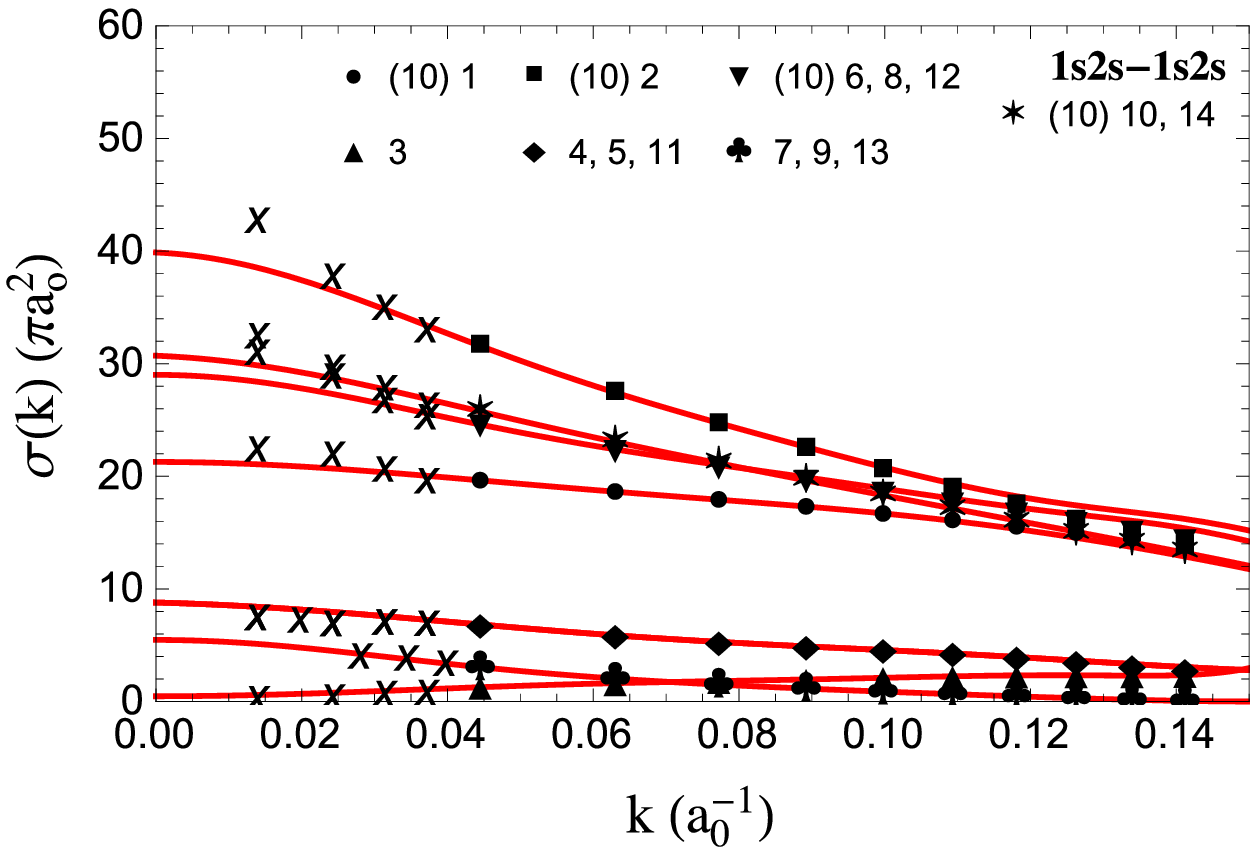}}
\caption{Cross-sections are given on a log scale for the (a) $1s1s\rightarrow1s1s$, (b) $1s2s\rightarrow1s1s$, and (c) $1s2s\rightarrow1s2s$ elastic and inelastic channels. The lines in each figure represent a curve-fit to the numerical results. The numbers by the point markers indicate row number in Table \ref{table:Spinprojectioncrosssections}. The $()$ labeling is a multiplying factor. Crosses are numerical results calculated at lower energies not used in the fitting procedure.}
\label{fig:crosssections}
\end{figure}

Experimental efforts to understand and manipulate Ps gases can be enhanced if one knows the Ps collision cross-sections. Cross-sections for different collision channels provide useful information about elastic and inelastic collision rates. For this reason, calculations for partial cross-sections for Ps collisions in uncoupled spin states are shown graphically in Fig. \ref{fig:crosssections} and numerically in Table \ref{table:Spinprojectioncrosssections}. 

Figures \ref{fig:crosssections} (a)--(c) show $k$-dependent partial cross-sections for in and out states of Ps atoms in specified uncoupled spin states for $1s1s\rightarrow1s1s$, $1s2s\rightarrow1s1s$, and $1s2s\rightarrow1s2s$ elastic and inelastic energy states, respectively. The wavenumber is that with respect to the $1s$--$2s$ energy threshold. Each curve is labeled by a number corresponding to a row in Table \ref{table:Spinprojectioncrosssections}, which indicates the corresponding collision channel. The $()$ labeling in Fig. \ref{fig:crosssections} (a) and (c) indicates a multiplying factor when reading the figure for the corresponding row number(s). Data represented as crosses are numerical results computed at smaller scattering energies not used in the fitting procedure. The discrepancies between the data at smaller energies and the fits can be attributed to errors in extracting infinite $R$ values of K-matrix elements as described at the end of section \ref{sec:numanalysis}.

Figure \ref{fig:crosssections} and Table \ref{table:Spinprojectioncrosssections} show interesting features that merit discussion. One feature is the behavior of the inelastic cross-section for the $1s2s\rightarrow1s1s$ process(shown in Fig. \ref{fig:crosssections} (b)). Near threshold, the inelastic cross-section is proportional to $k^{-1}$, which is the expected behavior for inelastic processes in accordance with the Wigner threshold law. This behavior is evident by the agreement of the fit to the numerical results and the extra points calculated over a smaller energy range, represented by crosses. The $1s1s\rightarrow1s2s$ inelastic cross-section goes to zero(linearly in $k$) at the $1s$--$2s$ threshold, thus is not shown.

Another feature to note is the spin-flip process from triplet to singlet states. This process occurs when the initial states are oppositely spin-aligned with $|M_{S}|=1$ and also for $M_{S}=0$. For the elastic $1s1s\rightarrow1s1s$ and $1s2s\rightarrow1s2s$ channels, the polarized cross-sections for spin collisions $S_{13}^{'}S_{24}^{'}\rightarrow S_{13}S_{24}=11\rightarrow 00$ with $M_{S}^{'}=\pm 1$ are non-negligible with values at the 1$s$--2$s$ threshold of approximately $1.68\pi a_{0}^{2}$, and $8.8 \pi a_{0}^{2}$, respectively. These cross-sections indicate the collision rate of triplet to singlet spin conversion is appreciable and would result in a shorter lifetime of a Ps(1$s$)-Ps(2$s$) gas due to the short lifetime of singlet Ps.

Another important feature is there exists no spin-flip channel for collisions of triplet Ps atoms in the same spin-stretched state. This is due to the fact that the total spin projection quantum number, $\mathrm{M_{S_{tot}}}$, is a conserved quantity during the collision process. A long-lived Ps gas can be obtained using these spin-stretched states with the effect of annihilation on the lifetime of the gas governed by the long-lived $s$-states of triplet Ps. Therefore, in a gas of unpolarized triplet Ps, there will be quenching of oppositely aligned spins, leading to a shorter lifetime then a gas of spin-aligned triplet Ps. 

\begin{table}[ht]
\caption{Zero-energy cross-sections for different spin and spin projection quantum numbers for Ps atom-atom collisions. Processes not shown are zero. All cross-sections are in atomic units of $\pi a_0^2$. For the $1s2s\rightarrow1s1s$ inelastic channel, $k\sigma$ (in units of $\pi a_{0}$) is given.}
\centering
\label{table:Spinprojectioncrosssections}
\resizebox{1\columnwidth}{!}{\begin{tabular}{l|l|lll}
    $\#$ & $S'_{13}M'_{13},S'_{24}M'_{24} \rightarrow S_{13}M_{13},S_{24}M_{24}$ & $\sigma_{1s1s\leftrightarrow1s1s}$ & $k\sigma_{1s2s\rightarrow1s1s}$ & $\sigma_{1s2s\leftrightarrow1s2s}$\\
    \hline
    $1$ & $(1,\pm1),(1,\pm1)\rightarrow(1,\pm1),(1,\pm1)$  & $21.1$ & $0.07$ & $213$ \\ 
    $2$ & $(1,\pm1),(1,\mp1)\rightarrow(1,\pm1),(1,\mp1)$  & $9.89$ & $0.56$ & $388$ \\
    $3$ & $(1,\pm1),(1,\mp1)\rightarrow(1,\mp1),(1,\pm1)$  & $9.89$ & $0.56$ & $0.47$ \\
    $4$ & $(1,\mp1),(1,\pm1)\rightarrow(1,0),(1,0)$ & $1.68$ & $0.43$ & $8.8$\\
    $5$ & $(1,\mp1),(1,\pm1)\rightarrow(0,0),(0,0)$  & $1.68$ & $0.43$ & $8.8$ \\
    $6$ & $(1,\pm1),(1,0)\rightarrow(1,\pm1),(1,0)$  & $21.1$ & $0.04$ & $290$ \\
    $7$ & $(1,\pm1),(1,0)\rightarrow(1,0),(1, \pm1)$  & $21.1$ & $0.04$ & $5.95$ \\
    $8$ & $(1,\pm1),(0,0)\rightarrow(1,\pm1),(0,0)$  & $10.6$ & $0.04$ & $290$ \\
    $9$ & $(1,\pm1),(0,0)\rightarrow(0,0),(1,\pm1)$  & $10.6$ & $0.04$ & $5.95$ \\
    $10$ & $(1,0),(1,0)\rightarrow(1,0),(1,0)$  & $14.7$ & $0.38$ & $307$ \\
    $11$ & $(1,0),(1,0)\rightarrow(0,0),(0,0)$  & $0.84$ & $0.21$ & $8.8$ \\
    $12$ & $(1,0),(0,0)\rightarrow(1,0),(0,0)$  & $10.6$ & $0.04$ & $290$ \\
    $13$ & $(1,0),(0,0)\rightarrow(0,0),(1,0)$  & $10.6$ & $0.04$ & $5.95$ \\
    $14$ & $(0,0),(0,0)\rightarrow(0,0),(0,0)$  & $14.7$ & $0.38$ & $307$\\\hline
\end{tabular}}
\end{table}

With the tabulated values for partial cross-sections in Table \ref{table:Spinprojectioncrosssections}, the spin-averaged cross-sections for polarized and unpolarized initial and final spin states can be computed. Formulas for spin-averaged cross-sections for different initial and final states are provided in Eqs. \ref{eq:spinavgcrosssections} (a)--(c) \cite{ScattTheory}.
\begin{subequations}
\begin{center}
\end{center}
\begin{equation}
    \sigma_{\substack{(S_{1i},M_{1i})\\(S_{2i},M_{2i})}\rightarrow S_{1f},S_{2f}}^{p\rightarrow u}=\sum_{M_{1f},M_{2f}}\sigma_{\substack{(S_{1i},M_{1i})\\(S_{2i},M_{2i})}\rightarrow\substack{(S_{1f},M_{1f})\\(S_{2f},M_{2f})}}
\end{equation}
\begin{center}
\end{center}
\begin{equation}
    \sigma_{S_{1i},S_{2i}\rightarrow \substack{(S_{1f},M_{1f})\\(S_{2f},M_{2f})}}^{u\rightarrow p}=\frac{1}{\alpha_i}\sum_{M_{1i},M_{2i}}\sigma_{\substack{(S_{1i},M_{1i})\\(S_{2i},M_{2i})}\rightarrow\substack{(S_{1f},M_{1f})\\(S_{2f},M_{2f})}}
\end{equation}
\begin{center}
\end{center}
\begin{equation}
    \sigma_{S_{1i},S_{2i}\rightarrow S_{1f},S_{2f}}^{u\rightarrow u}=\frac{1}{\alpha_i}\sum_{M_{1i},M_{2i}}\sum_{M_{1f},M_{2f}}\sigma_{\substack{(S_{1i},M_{1i})\\(S_{2i},M_{2i})}\rightarrow\substack{(S_{1f},M_{1f})\\(S_{2f},M_{2f})}}
\end{equation}
\label{eq:spinavgcrosssections}
\end{subequations}

\noindent The labels $p\rightarrow u$, $u\rightarrow p$, and $u\rightarrow u$ indicate initial and final spin states as either polarized(p) or unpolarized(u). The indices 1 and 2 are particle labels for two-particle collisions(Eqs. \ref{eq:spinavgcrosssections} (a)--(c) can be written similarly for an arbitrary number of spins). The prefactor $\alpha_i$ in Eqs. \ref{eq:spinavgcrosssections} (b) and (c) is defined as $\alpha_i=(2S_{1i}+1)(2S_{2i}+1)$.
\section{Conclusions}
\label{sec:conclusions}
This paper presents a study of the $e^-e^-e^+e^+$ system with total angular momentum $L$=$\mathrm{0}$ and positive parity that treats the elastic and inelastic scattering properties of the Ps(1$s$)-Ps(2$s$) dimer-dimer breakup channel. A motivation for our exploration of this system is to determine the triplet-triplet, singlet-singlet, and singlet-triplet scattering lengths that would be needed to quantify the clock-shift frequency for Ps density characterization in an atomic gas and BEC. The scattering lengths for the TT, SS, and ST spin configurations are calculated to be $a_{\mathrm{TT}}=$~$7.3(2)a_0-i0.02(1)a_0$, $a_{\mathrm{SS}}=$~$13.2(2)a_0-i0.9(2)a_0$, and $a_{\mathrm{ST}}=a_{\mathrm{TS}}=$~$9.7(2)a_0$ respectively.

Spin re-coupling is implemented to describe coupled positronium spin states for a given total spin in terms of the electron-electron and positron-positron spin representation. As a result, the scattering lengths describing the collisions of positronium atoms in the $1s2s$ elastic channel are obtained for Ps in uncoupled spin states. For a gas of spin-polarized triplet Ps, the clock-shift frequency per unit density is estimated to be $\frac{\Delta \nu_{1s-2s}}{n}=10.0(2)\times10^{-8} cm{^3}$-$Hz$. 

Cross-sections are quantified for different spin collisions relative to the $1s2s$ threshold energy. A brief description is provided on how to calculate scattering cross-sections for Ps collisions prepared and measured in polarized and unpolarized spin states.

\section*{Acknowledgements}
This work was supported in part by the U.S. Department of Energy, Office of Science, DE-SC0010545. We would also like to thank David B. Cassidy for fruitful discussions and suggestions.

\bibliography{bibliography}
\end{document}